\documentclass[runningheads]{llncs}
\usepackage{graphicx}
\usepackage{subfigure}
\usepackage{multirow}
\usepackage{hyperref}
\usepackage[title]{appendix}
\usepackage{comment}

\usepackage{xcolor}
\usepackage{enumitem}
\begin{document}
%
\title{XCAT-GAN for Synthesizing 3D Consistent Labeled Cardiac MR Images on Anatomically Variable XCAT Phantoms}
\titlerunning{XCAT-GAN: A Novel Approach for Data Augmentation}
%
%
\makeatletter
\newcommand{\printfnsymbol}[1]{%
  \textsuperscript{\@fnsymbol{#1}}%
}
\makeatother

\author{Sina Amirrajab\inst{1}\thanks{Contributed equally} 
\and
Samaneh Abbasi-Sureshjani\inst{1}\printfnsymbol{1} 
\and
Yasmina Al Khalil\inst{1}\printfnsymbol{1} 
\and
Cristian Lorenz\inst{2} 
\and
Juergen Weese\inst{2} 
\and
Josien Pluim\inst{1} 
\and
Marcel Breeuwer\inst{1,3} 
}
\authorrunning{S. Amirrajab et al.}
%
\institute{Eindhoven University of Technology, Eindhoven, The Netherlands \email{\{s.amirrajab,s.abbasi,y.al.khalil,j.pluim,m.breeuwer\}@tue.nl} \and
Philips Research Laboratories, Hamburg, Germany
\email{\{cristian.lorenz,juergen.weese\}@philips.com}
\and
Philips Healthcare, MR R\&D - Clinical Science, Best, The Netherlands
}


\maketitle              
\begin{abstract}

\sloppy
Generative adversarial networks (GANs) have provided promising data enrichment solutions by synthesizing high-fidelity images. However, generating large sets of labeled images with new anatomical variations remains unexplored. We propose a novel method for synthesizing cardiac magnetic resonance (CMR) images on a population of virtual subjects with a large anatomical variation, introduced using the 4D eXtended Cardiac and Torso (XCAT) computerized human phantom. We investigate two conditional image synthesis approaches grounded on a semantically-consistent mask-guided image generation technique: 4-class and 8-class XCAT-GANs. The 4-class technique relies on only the annotations of the heart; while the 8-class technique employs a predicted multi-tissue label map of the heart-surrounding organs and provides better guidance for our conditional image synthesis. For both techniques, we train our conditional XCAT-GAN with real images paired with corresponding labels and subsequently at the inference time, we substitute the labels with the XCAT derived ones. Therefore, the trained network accurately transfers the tissue-specific textures to the new label maps. By creating 33 virtual subjects of synthetic CMR images at the end-diastolic and end-systolic phases, we evaluate the usefulness of such data in the downstream cardiac cavity segmentation task under different augmentation strategies. Results demonstrate that even with only 20\% of real images (40 volumes) seen during training, segmentation performance is retained with the addition of synthetic CMR images. Moreover, the improvement in utilizing synthetic images for augmenting the real data is evident through the reduction of Hausdorff distance up to 28\% and an increase in the Dice score up to 5\%, indicating a higher similarity to the ground truth in all dimensions.

\end{abstract}
%
%
%
\section{Introduction}
The medical image analysis community is suffering from limited annotated data, constrained sharing policies, and imbalanced samples, which directly lead to limitations in the adoption of novel and reliable deep learning methods. This is particularly reflected in clinical practice, where deep learning models lack robustness and generalizability to the diverse patient population, as well as variations in imaging protocols, acquisition parameters and manufacturing specifications of imaging devices \cite{2018Yasaska}. 
The image synthesis techniques by Generative Adversarial Networks (GANs)~\cite{GANs_Goodfellow2014}  have gained lots of attention as a potential approach to address these problems by augmenting and balancing the data in various domains \cite{ReviewGAN,kazeminia2018gans}. 
For instance, leveraging the synthetic data obtained by domain translation techniques led to significant performance improvements in the application of cardiac cavity segmentation \cite{2019Chen_Munit_Style,CardiacChartsias2017,CardiacBaumgartner2019}. However, in most domain translation approaches, the variations in the synthetic images are constrained by the source domain i.e., either the input label maps or the images in the source domain have limited anatomical variations. Thus, generating new anatomically-plausible instances is not possible.
\subsection{Related Work}

Separating the style (often called domain-specific features) from the shape (often called domain-invariant features) has enabled various domain translation techniques~\cite{huang2018munit,park2019SPADE,wang2018pix2pixHD}, which can synthesize high-resolution photo-realistic images with detailed textures from semantic label maps using conditional GANs.
A recent work by~\cite{LaparoscopicSynthesis2019} showed the effectiveness of translating the simulated to realistic images for the task of liver segmentation in laparoscopic images using a conditional unpaired image-to-image translation approach~\cite{huang2018munit}. 
The works by~\cite{2019Chen_Munit_Style,StyleCardiacSeg_MICCAI19} used the same technique to combine the styles from domains with few or no labels with the contents of another domains with more labels and use them in adapting the segmentation networks. Even though such techniques provide flexibility by using unpaired sets, paired image-to-image translation techniques~\cite{wang2018pix2pixHD,park2019SPADE} provide more control over image generation. They enable generating semantically consistent labeled synthetic images, which are more useful for medical image analysis (such as the work by~\cite{CT_aug_SPIE2019}). However, the only possible variation in synthetic data can be created by learning different styles (often via Variational Auto-Encoders (VAE) \cite{VAEsWelling}) and then mixing them with the existing shape information provided by the annotated label maps with limited anatomical variations. Although it is possible to model and create new shape deformations statistically~\cite{StatisticalShapeVariations2019} or via VAEs~\cite{factorized_ETH}, such models are not necessarily anatomically plausible and structurally accurate.

\subsection{Contributions}

In this paper, we address the above-mentioned challenges by proposing a novel framework for synthesizing a diverse labeled database of 3D consistent Cardiac Magnetic Resonance (CMR) images on anatomically-plausible XCAT derived labels.
%
The new framework, called XCAT-GAN, is 
tailored to transfer the modality-specific MR image characteristics to new anatomical representations to generate a new realistic-looking synthetic CMR image database in two ways: 4-class and 8-class image synthesis. We rely on the eXtended CArdiac Torso (XCAT) \cite{segars20104d} phantoms as the source of the computerized human phantoms to generate 33 virtual subjects with various anatomical features. 
Our synthetic database comprises heterogeneous images of virtual subjects that are variable in terms of anatomical representation and diverse in terms of image appearance. Manual annotation is not required since the anatomical model used for image generation serves as the ground truth label map for synthetic images. Real annotated datasets are needed for training the network, but the existing annotations are limited to a few classes of the heart.
For improving the guidance in image generation, multi-tissue label maps of surrounding regions of the heart are obtained by leveraging a segmentation network trained on a set of simulated images~\cite{ISMRM2020SimXCAT}. 
Using these new label maps leads to a substantial improvement in the consistency of the synthesis particularly of the organs for which the annotations are not available. Finally, we demonstrate that the synthetic data can be used not only for augmenting the datasets but also for replacing the real images since they provide accurate and realistic shapes and appearances.
Compared to our concurrent image synthesis approach in \cite{abbasi-sureshjani2020d}, we achieve anatomical consistency in the third dimension by incorporating more labels for the organs visible in the background. Furthermore, we validate the utility of the synthetic images for a clinical task quantitatively, which has not been explored previously.
\section{Methodology}
\begin{figure}[h!]
\includegraphics[width=\textwidth]{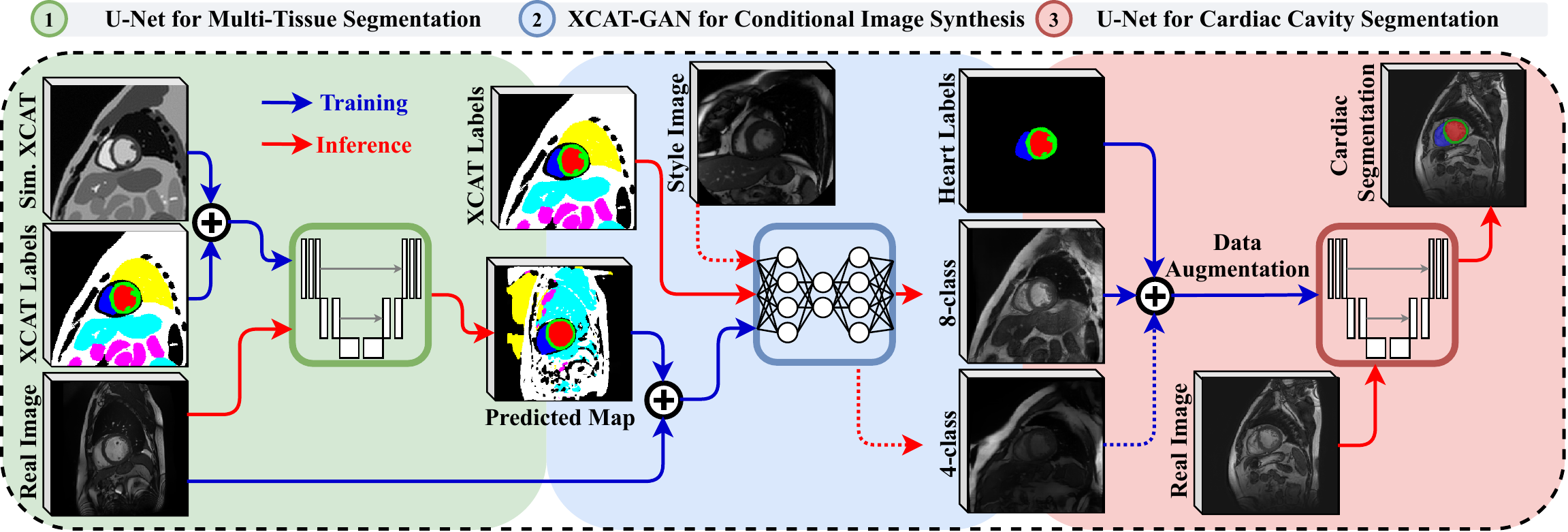}
\caption{An overview of the three networks used in our proposed approaches. 
The first U-Net is trained completely on the simulated CMR data with 8-class ground truth labels. The paired images and label maps (4 or 8 classes) are used to train the second network for  conditional image synthesis. At the inference, the XCAT derived labels are used for synthesizing 3D CMR images. New styles are also transferred to the synthetic images in the 4-class XCAT-GAN.  The synthetic data with its ground truth labels are evaluated by the third network in a supervised cardiac cavity segmentation task.} \label{fig:overview}
\end{figure}

We demonstrate a novel image synthesis approach with two different settings; for the first one we only use available ground truth labels for the heart (4-class), and in the second one we increase the number of labels (8-class) when training the XCAT-GAN. A general overview of the networks is illustrated in Fig.~\ref{fig:overview}. 

Three neural networks are used for image generation and evaluation.
Network 1 is a modified U-Net~\cite{Ronneberger} that predicts multi-tissue segmentation maps for a set of real images. This component is  only used for the 8-class image synthesis to provide more guidance for our XCAT-GAN. Network 2 is a conditional GAN architecture~\cite{park2019SPADE}, trained on paired real images and label maps (either 4 or 8-class maps).
After the training is done, it 
synthesizes images on the XCAT labels as the new anatomical representations. In the 4-class image synthesis additional images are used to add variations in the style of generated images at the inference time.
The new synthetic images with their corresponding cardiac labels are evaluated in different experiment settings by network 3 that is an adapted version of the 2D U-Net architecture \cite{Isensee2019nnUNetBT}.
All three networks are trained separately so that the output of the first network is used as the input of the second network in the training time and the output of the second network is used as the data augmentation for the third network. In the following sections, we explain the data used at each stage and elaborate on the particular strategies for training and inference times.

\subsection{Material}
We utilize four real CMR image datasets and one simulated CMR image dataset in our proposed framework. The simulated dataset contains 66 simulated CMR image volumes that are generated with MRXCAT analytical approach \cite{wissmann2014mrxcat} on variable XCAT phantoms~\cite{ISMRM2020SimXCAT} with anatomical variabilities such as heart size, location and orientation with respect to other organs. One sample of a simulated XCAT image and its corresponding label map is shown at the top left of Fig.~\ref{fig:overview}. The Automated Cardiac Diagnosis Challenge (ACDC) dataset~\cite{ACDC} is the main real dataset used for training the XCAT-GAN. The Sunnybrook Cardiac Data (SCD)~\cite{SCD} and York Cardiac MRI (York)~\cite{York} datasets are only used for adding style variations in the 4-class image synthesis at the inference time and are not seen during the training. Moreover, 156 internal clinical CMR (cCMR) volumes are used for our quantitative evaluation step\footnote{All data used in this study were obtained with the required approvals and patient consent.}. All datasets are pre-processed by sub-sampling them to $1.3\times1.3~mm$ in-plane resolution and taking a central crop of the images with $256\times256$ pixels. All intensity values are normalized to the range $[-1, 1]$. Data is augmented during the training by random scaling, rotations, elastic deformation, and mirroring. Further details about each dataset are available in Appendix A, Table 1.

\subsection{Network Details}
A brief explanation about each network is provided in this section. Further details about the training and architectures are available in Appendix B. 

\textbf{U-Net for Multi-Tissue Segmentation} (network 1) is used for obtaining a coarse multi-tissue segmentation map of the heart cavity and other background tissues in the 8-class image synthesis setting. 
We adopt a U-Net architecture~\cite{Ronneberger}, completely trained on the XCAT simulated dataset with its 8-class ground truth masks in a supervised manner. 
 In the inference time, the real MR image from the ACDC or cCMR dataset is fed to the trained network and a rough segmentation map is created. The heart labels are replaced with the manual annotations before feeding them to the next network, along with their corresponding real images.

\textbf{XCAT-GAN for Conditional Image Synthesis}~\cite{park2019SPADE} is a mask-guided image generation technique that employs spatially-adaptive denormalization layers resulting in semantically-consistent image synthesis. During the training, the network takes the (predicted) label maps of the ACDC dataset with their corresponding real images and learns their underlying modality-specific features for each class.   
Since the guidance in the 4-class synthesis case is limited to three heart classes and the background contains various organs, we provide more guidance to the network by using a style encoder in a VAE setting~\cite{park2019SPADE} without instance normalization layers that often lead to the removal of semantic information.  This helps in learning and transferring new styles and structures in the background. At the inference time, the label maps are replaced with the XCAT labels. Also, additional style images are used as the input of the style encoder. 

 \textbf{U-Net for Cardiac Cavity Segmentation} (network 3) is employed for segmentation of heart cavities in real cardiac data with the addition of labelled synthetic XCAT images in different augmentation settings. This network is a modified 2D U-Net, designed and optimized following the recommendations of the nnU-Net framework \cite{Isensee2019nnUNetBT}, as explained in Appendix B.

\subsection{Experiments and Results}

\textbf{Image Synthesis:}
For synthesizing the images, we utilize 66 XCAT volumes including both end-diastolic (ED) and end-systolic (ES) phases (33 subjects). For the 4-class case, we utilize 22 random volumes generated with three different styles from the York, SCD, and ACDC dataset, respectively; while for the 8-class case, we use 66 volumes synthesized only with the ACDC style. Throughout the whole experiments, we keep the number of 4-class and 8-class synthetic images the same (66 volumes) to have a fair comparison in our segmentation task.

\begin{figure}[t]
\centering
{
\subfigure[The synthetic image for a 4-class label map]{\label{fig:4-class}\includegraphics[width=\textwidth]{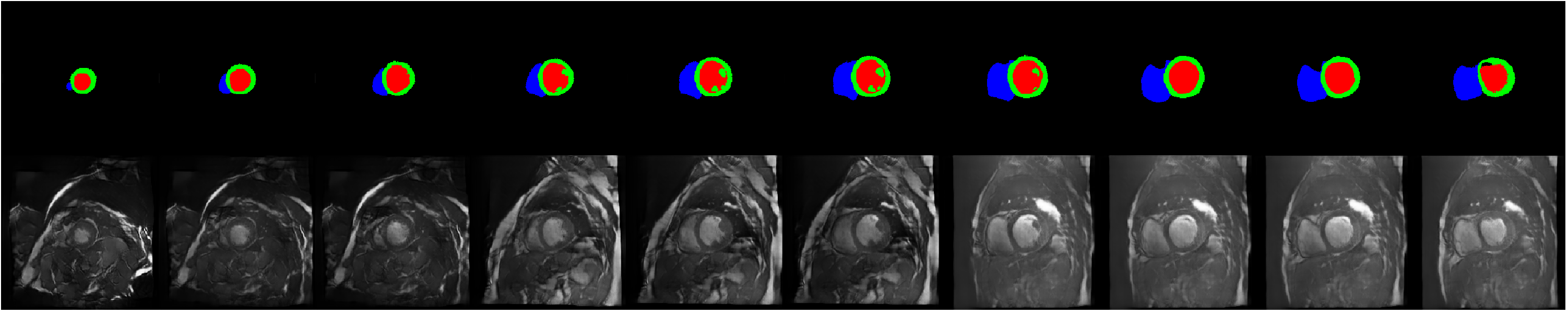}}

\subfigure[The synthetic image for an 8-class label map]{\label{fig:8-class}\includegraphics[width=\textwidth]{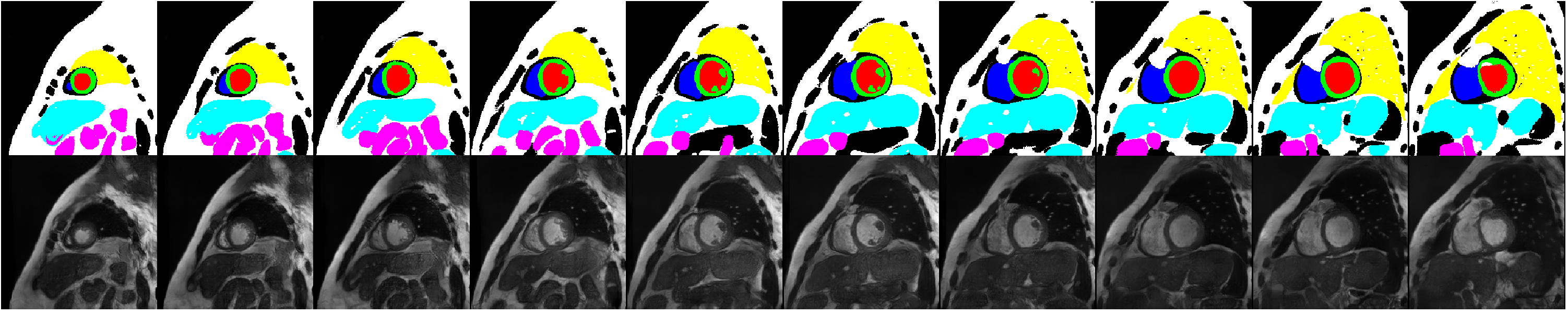}}
}
{\caption{Sample images generated by \subref{fig:4-class} 4 and \subref{fig:8-class} 8-class XCAT-GAN with their correspinding XCAT labels for 10 short axis slices from the apex to the base of the heart. 
} \label{fig:image synthesis}}
\end{figure}

Fig.~\ref{fig:image synthesis} depicts sample synthetic images generated by 4-class (with the ACDC style) and 8-class XCAT-GANs. 
Moreover, a sample 4-class synthetic image with the SCD style is depicted in Fig.~\ref{fig:overview}, between networks 2 and 3.
As seen in Fig.~\ref{fig:4-class}, in the background region, both the texture and the anatomical content including the surrounding organs and structures to the heart such as the fat layer of the outer layer of the body, lung and abdominal organs are generated well by the network. Although the synthetic texture for heart classes seems to be consistent and robust for different slices across the heart long axes view, the synthetic content for the background class changes. Considering the complexity of the anatomical content and style for the background class, this inconsistency comes as no surprise. To alleviate this, the 8-class XCAT-GAN is proposed. 

The 8-class XCAT-GAN uses the predictions by network 1.  A sample of a full predicted map is depicted in Fig.~\ref{fig:overview}, between the first and second networks. As seen in Fig.~\ref{fig:8-class}, 
as a result of utilizing more labels for training the XCAT-GAN, more local characteristic for each visible structure in the image are learned. Subsequently at the inference time, when we feed XCAT labels with eight classes, the synthetic images are consistent in the third direction. 
Two 3D visualizations of the synthetic images for the ED and ES phases are available at \url{https://bit.ly/2wo9KOE} and \url{https://bit.ly/3aeuoiC}, respectively.

\textbf{Cardiac Cavity Segmentation:}
To quantitatively evaluate the usefulness of the synthetic cardiac images, we combine them with real images to train a network for cardiac cavity segmentation task. As a baseline, we first train our network on the existing ACDC and cCMR images for three classes: left ventricular myocardium (MYO), left ventricle blood pool (LV) and right ventricle blood pool (RV). In the first set of experiments, we evaluate the effect of adding synthetic images to the real MR data during training (\emph{Augmentation}). Other experiments focus on gradually reducing real MR data in the training set, while the number of synthetic images is kept constant (\emph{Real Data Reduction}). The aim is to evaluate whether the synthetic data are realistic enough to replace real MR data.
\begin{figure}[h!]
\includegraphics[width=0.73\textwidth]{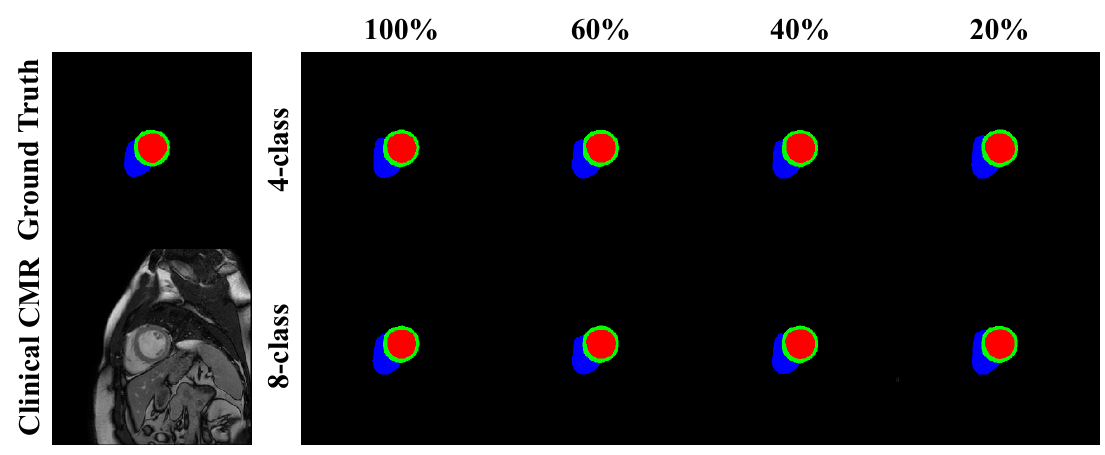}
\centering
\caption{Sample segmentation results of the real data reduction experiment. Percentages indicate the ratio of used real data relative to the baseline experiment (200 volumes).%
}
\label{fig:reduction}
\end{figure}
We use the same data for inference in all experiments, namely the subsets of both cCMR and ACDC datasets. Reference Dice scores (DSC) achieved on the ACDC dataset are $0.968$, $0.946$ and $0.902$ for the ED phase and $0.931$, $0.889$ and $0.919$ for the ES phase on LV, RV, and MYO classes, respectively \cite{ACDC}.
A configuration summary of all experiments and results can be found in Table~\ref{tab:experiments2}.\begin{table}[b!]
\centering
\caption{Segmentation results for augmentation and real data reduction experiments. 
}
\label{tab:experiments2}
\resizebox{\linewidth}{!}{%
\begin{tabular}{c|cc|cc|cc|cc|cc|cc|cc|cc} 
\hline
Exp.                                                         & \multicolumn{4}{c|}{Training Set}                                & \multicolumn{6}{c|}{Clinical MR Test Set (n=50)}                                                       & \multicolumn{6}{c}{ACDC Test Set (n=50)}                                                                \\ 
\hline
\multicolumn{1}{c|}{\multirow{8}{*}{\rotatebox{90}{Augmentation}}} & \multicolumn{2}{c|}{Real Set} & \multicolumn{2}{c|}{Synthetic Set} & \multicolumn{2}{c|}{LV}          & \multicolumn{2}{c|}{MYO}         & \multicolumn{2}{c|}{RV}          & \multicolumn{2}{c|}{LV}          & \multicolumn{2}{c|}{MYO}         & \multicolumn{2}{c}{RV}            \\
\multicolumn{1}{c|}{}                                        & Name                 & \# Vol.     & Name         & \# Vol.             & DSC            & HD              & DSC            & HD              & DSC            & HD              & DSC            & HD              & DSC            & HD              & DSC            & HD               \\ 
\cline{2-17}
\multicolumn{1}{c|}{}                                        & cCMR         & 100   & -            & -                  & 0.91           & 10.11           & 0.84           & 13.74           & 0.88           & 11.73           & 0.86           & \textbf{12.57}  & 0.80           & 14.61           & 0.87           & 22.73            \\
\multicolumn{1}{c|}{}                                        & cCMR         & 100   &  4-class           & 66                  & 0.93          & 10.73         & 0.86           & 13.98           & 0.89           & 9.94          & 0.90           & 14.25  & 0.90           & 15.54           & 0.86           & 16.72            \\
\multicolumn{1}{c|}{}                                        & cCMR         & 100   & 8-class  & 66                 & \textbf{0.94}  & \textbf{9.84}   & \textbf{0.86}  & \textbf{8.06}   & \textbf{0.89}  & \textbf{8.08}   & \textbf{0.91}  & 12.98           & \textbf{0.91}  & \textbf{13.40}  & \textbf{0.87}           & \textbf{11.87}   \\ 
\cline{2-17}
\multicolumn{1}{c|}{}                                        & ACDC                 & 100   & -            & -                  & 0.88           & \textbf{20.46}  & 0.81           & 31.26           & 0.82           & 26.28           & 0.94           & 11.21           & 0.90           & 12.94           & 0.92           & 14.14            \\
\multicolumn{1}{c|}{}                                        & ACDC                 & 100   & 4-class  & 66                 & 0.88  & 33.50           & 0.81  & 41.39  & 0.84  & 20.59  & 0.95 & 11.81   & 0.91  & 12.25  & \textbf{0.94}  & 11.87   \\ 
\multicolumn{1}{c|}{}                                        & ACDC                 & 100   & 8-class  & 66                 & \textbf{0.89}  & 21.84           & \textbf{0.82}  & \textbf{25.71}  & \textbf{0.84}  & \textbf{18.89}  & \textbf{0.96}  & \textbf{9.73}   & \textbf{0.92}  & \textbf{11.91}  & 0.93  & \textbf{10.67}   \\ 
\hline\hline
\multirow{12}{*}{\rotatebox{90}{Real Data Reduction}}              & cCMR+ACDC  & 200   & -            & -                  & 0.93           & \textbf{9.84}   & 0.85           & 13.79           & 0.89           & 10.63           & 0.95           & 8.71            & 0.90           & 11.33           & 0.91           & 12.86            \\
                                                             & cCMR+ACDC  & 200   & 4-class & 66                 & 0.93           & 12.15           & 0.85           & 15.27           & 0.89           & 12.93           & 0.95           & 9.51            & 0.91           & 15.99           & 0.92           & 16.63            \\
                                                             & cCMR +ACDC & 200   & 8-class & 66                 & 0.93           & 9.89            & \textbf{0.86}  & \textbf{13.22}  & 0.89           & \textbf{9.43}   & 0.95           & \textbf{7.99}   & \textbf{0.92}  & \textbf{9.07}   & \textbf{0.92}  & \textbf{10.32}   \\ 
\cline{2-17}
                                                             & cCMR+ACDC  & 120   & -            & -                  & 0.90           & 12.11           & 0.83           & 15.73           & 0.86           & 14.39           & 0.94           & 9.14            & 0.89           & 15.08           & 0.88           & 16.35            \\
                                                             & cCMR+ACDC  & 120   & 4-class  & 66                 & 0.92           & 10.64           & 0.85           & 14.27           & 0.87           & 12.20           & 0.95           & 10.12           & 0.90           & 12.77           & 0.92           & 12.21            \\
                                                             & cCMR+ACDC  & 120   & 8-class  & 66                 & \textbf{0.92}  & \textbf{9.98}   & \textbf{0.85}  & \textbf{13.27}  & \textbf{0.89}  & \textbf{10.11}  & \textbf{0.95}  & \textbf{8.89}   & \textbf{0.91}  & \textbf{11.31}  & \textbf{0.93}  & \textbf{12.08}   \\ 
\cline{2-17}
                                                             & cCMR+ACDC  & 80    & -            & -                  & 0.87           & 17.22           & 0.82           & \textbf{18.01}  & 0.86           & 15.76           & 0.92           & 16.51           & 0.87           & 17.22           & 0.87           & 18.11            \\
                                                             & cCMR+ACDC  & 80    & 4-class & 66                 & 0.92           & 21.34           & 0.85           & 24.77           & 0.88           & 18.77           & 0.93           & 16.75           & 0.89           & 19.57           & 0.89           & 19.15            \\
                                                             & cCMR+ACDC  & 80    & 8-class & 66                 & \textbf{0.92}  & \textbf{14.67}  & \textbf{0.85}  & 18.82           & \textbf{0.88}  & \textbf{11.72}  & \textbf{0.94}  & \textbf{11.30}  & \textbf{0.91}  & \textbf{12.79}  & \textbf{0.90}  & \textbf{14.76}   \\ 
\cline{2-17}
                                                             & cCMR+ACDC  & 40    & -            & -                  & 0.85           & 21.13           & 0.79           & 22.69           & 0.83           & 18.91           & 0.89           & 19.74           & 0.85           & 19.11           & 0.85           & 20.05            \\
                                                             & cCMR+ACDC  & 40    & 4-class & 66                 & 0.90           & 26.29           & 0.82           & 38.73           & 0.85           & 24.42           & 0.92           & 22.79           & 0.89           & 27.13           & 0.88           & 28.33            \\
                                                             & cCMR+ACDC  & 40    & 8-class  & 66                 & \textbf{0.91}  & \textbf{16.28}  & \textbf{0.84}  & \textbf{19.04}  & \textbf{0.87}  & \textbf{13.75}  & \textbf{0.94}  & \textbf{15.49}  & \textbf{0.90}  & \textbf{15.76}  & \textbf{0.90}  & \textbf{17.38}   \\
\hline
\end{tabular}
}
\end{table} 

The \emph{augmentation} experiments show that utilizing the synthetic data improves the segmentation performance of the network trained on cCMR and ACDC images in terms of the mean DSC and Hausdorff distance (HD) for most cases. Moreover, utilizing 8-class synthetic data improves the HD significantly across all experiments.

The \emph{real data reduction} experiments indicate that segmentation performance is retained in cases when synthetic XCAT data is added to fewer real data in the training set compared to the baseline model (trained with 200 cCMR and ACDC images). Since the number of total volumes used for training is decreased in each experiment, we expect the performance to decrease accordingly. However, we do not observe a significant drop in performance as we introduce new shape priors through the addition of synthetic images. This holds even in cases when only $20\%$ of the baseline MR real images are used in the training set. Some sample segmentation results on the cCMR dataset for this experiment are also depicted in Fig.~\ref{fig:reduction}, which shows the segmentation maps remain accurate when replacing the real data with  synthetic ones. 

\section{Discussion and Conclusion}
In this paper, we propose a novel framework for generating a database of synthetic 3D consistent labeled CMR images for medical data augmentation. Our XCAT-GAN combines the anatomically accurate XCAT labels with the learned modality-specific image characteristics to synthesize plausible images.
In the 4-class image synthesis setting, XCAT-GAN can learn both the texture of the heart and the underlying content information for the background. Furthermore, by incorporating more labels for the surrounding regions of the heart in the 8-class synthesis, we achieve more consistency and robustness in synthesising 3D image volumes, despite the fact that we used 2D networks for all of our experiments. 

We evaluate the usefulness of this synthetic database in data augmentation for the supervised cardiac cavity segmentation task. For both 4-class and 8-class synthesis, it is demonstrated that the segmentation performance is retained even when the network is trained on fewer real data i.e., a significant number of real MR images (up to $80\%$) can be replaced by synthetic images during training, which is a promising step towards addressing the lack of available medical data. While the DSC is slightly improved in the 8-class augmentation compared to 4-class, the false positives of the background tissues are significantly reduced due to 3D consistency achieved by providing more labels in the process of synthesis. This is also demonstrated by obtaining a smaller HD. However, such results could also be the effect of training and evaluation with a 2D segmentation network only. Thus, future work includes utilizing the 8-class images in a 3D segmentation setting. Nevertheless, images synthesized with the method proposed in this paper can supplement or be utilized instead of real data.


Our rough segmentation maps by the first network are not perfect due to the gap between the simulated images (used for training) and real images (used for testing). This leads to false conditioning in the XCAT-GAN in some regions. It is possible to improve the quality of the segmentation maps, by improving the segmentation network or using the domain adaptation techniques such as~\cite{ACE2019}. The other option is to use other labeled real sets with labels covering all regions rather than only the heart classes (if available). Moreover, the style or modality-specific feature mentioned in this paper is learned from the real data with cine MR contrast. Since all the datasets we used have very similar styles, the main advantage of the VAE setup for the 4-class image synthesis is to introduce more control over the surrounding regions. Rather than memorizing and generating the same background, it learns to transfer some structures from the style image as well, which is why the 4-class and 8-class synthetic images lead to comparable results in our experiments.
Transferring the style of other MR modalities such as T1-weighted and late gadolinium enhancement conditioned on each anatomical label class is considered as a future direction.  Finally, it is worth mentioning that it is possible to extend the image synthesis to 4D by using the  parameterized  motion  model  of  the  XCAT heart and generating 3D+t CMR images.

\subsection*{Acknowledgments}This research is a part of the openGTN project, supported by the European Union in the Marie Curie Innovative Training Networks (ITN) fellowship program under project No. 764465.


%

\bibliographystyle{splncs04}
\bibliography{paper2194}

\begin{thebibliography}{10}
\providecommand{\url}[1]{\texttt{#1}}
\providecommand{\urlprefix}{URL }
\providecommand{\doi}[1]{https://doi.org/#1}

\bibitem{abbasi-sureshjani2020d}
Abbasi-Sureshjani, S., Amirrajab, S., Lorenz, C., Weese, J., Pluim, J.,
  Breeuwer, M.: {4D} semantic cardiac magnetic resonance image synthesis on
  {XCAT} anatomical model. In: Medical Imaging with Deep Learning (2020)

\bibitem{ISMRM2020SimXCAT}
Amirrajab, S., Al~Khalil, Y., Lorenz, C., Weese, J., Breeuwer, M.: Towards
  generating realistic and hetrogeneous cardiac magnetic resonance simulated
  image database for deep learning based image segmentation algorithms.
  Proceedings of the 12th Annual Meeting ISMRM Benelux Chapter 2020; P-077.
  (2020)

\bibitem{York}
Andreopoulos, A., Tsotsos, J.K.: Efficient and generalizable statistical models
  of shape and appearance for analysis of cardiac {MRI}. Medical Image Analysis
   \textbf{12}(3),  335 -- 357 (2008)

\bibitem{ACDC}
Bernard, O., Lalande, A., Zotti, C., Cervenansky, F., et~al.: Deep learning
  techniques for automatic {MRI} cardiac multi-structures segmentation and
  diagnosis: Is the problem solved? IEEE Transactions on Medical Imaging
  \textbf{37}(11),  2514--2525 (2018)

\bibitem{CardiacBaumgartner2019}
Chaitanya, K., Karani, N., Baumgartner, C.F., Becker, A., Donati, O.,
  Konukoglu, E.: Semi-supervised and task-driven data augmentation. In: Chung,
  A.C.S., Gee, J.C., Yushkevich, P.A., Bao, S. (eds.) Information Processing in
  Medical Imaging. pp. 29--41. Springer International Publishing, Cham (2019)

\bibitem{CardiacChartsias2017}
Chartsias, A., Joyce, T., Dharmakumar, R., Tsaftaris, S.A.: Adversarial image
  synthesis for unpaired multi-modal cardiac data. In: Tsaftaris, S.A., Gooya,
  A., Frangi, A.F., Prince, J.L. (eds.) Simulation and Synthesis in Medical
  Imaging. pp. 3--13. Springer International Publishing, Cham (2017)

\bibitem{2019Chen_Munit_Style}
{Chen}, C., {Ouyang}, C., {Tarroni}, G., {Schlemper}, J., {Qiu}, H., {Bai}, W.,
  {Rueckert}, D.: Unsupervised multi-modal style transfer for cardiac {MR}
  segmentation. arXiv e-prints arXiv:1908.07344 (Aug 2019)

\bibitem{StatisticalShapeVariations2019}
Corral~Acero, J., Zacur, E., Xu, H., Ariga, R., Bueno-Orovio, A., Lamata, P.,
  Grau, V.: {SMOD} - data augmentation based on statistical models of
  deformation to enhance segmentation in {2D} cine cardiac {MRI}. In:
  Coudi{\`e}re, Y., Ozenne, V., Vigmond, E., Zemzemi, N. (eds.) Functional
  Imaging and Modeling of the Heart. pp. 361--369. Springer International
  Publishing, Cham (2019)

\bibitem{GANs_Goodfellow2014}
Goodfellow, I., Pouget-Abadie, J., Mirza, M., Xu, B., et~al.: Generative
  adversarial nets. In: Ghahramani, Z., Welling, M., Cortes, C., Lawrence,
  N.D., Weinberger, K.Q. (eds.) Advances in Neural Information Processing
  Systems 27, pp. 2672--2680. Curran Associates, Inc. (2014)

\bibitem{huang2018munit}
Huang, X., Liu, M.Y., Belongie, S., Kautz, J.: Multimodal unsupervised
  image-to-image translation. In: Ferrari, V., et~al. (eds.) Computer Vision --
  ECCV 2018. pp. 179--196. Springer International Publishing, Cham (2018)

\bibitem{Isensee2019nnUNetBT}
Isensee, F., Petersen, J., Kohl, S.A.A., J{\"a}ger, P.F., Maier-Hein, K.:
  {nnU-Net}: Breaking the spell on successful medical image segmentation. ArXiv
   \textbf{abs/1904.08128} (2019)

\bibitem{factorized_ETH}
Joyce, T., Kozerke, S.: {3D} medical image synthesis by factorised
  representation and deformable model learning. In: Burgos, N., et~al. (eds.)
  Simulation and Synthesis in Medical Imaging. pp. 110--119. Springer
  International Publishing, Cham (2019)

\bibitem{kazeminia2018gans}
Kazeminia, S., Baur, C., Kuijper, A., van Ginneken, B., Navab, N., Albarqouni,
  S., Mukhopadhyay, A.: Gans for medical image analysis (2018)

\bibitem{VAEsWelling}
{Kingma}, D.P., {Welling}, M.: Auto-encoding variational bayes. arXiv e-prints
  arXiv:1312.6114 (Dec 2013)

\bibitem{StyleCardiacSeg_MICCAI19}
Ma, C., Ji, Z., Gao, M.: Neural style transfer improves {3D} cardiovascular
  {MR} image segmentation on inconsistent data. In: Shen, D., et~al. (eds.)
  Medical Image Computing and Computer Assisted Intervention -- MICCAI 2019.
  pp. 128--136. Springer International Publishing, Cham (2019)

\bibitem{park2019SPADE}
Park, T., Liu, M.Y., Wang, T.C., Zhu, J.Y.: Semantic image synthesis with
  spatially-adaptive normalization. In: 2019 IEEE/CVF Conference on Computer
  Vision and Pattern Recognition (CVPR). pp. 2332--2341. IEEE Computer Society,
  Los Alamitos, CA, USA (jun 2019)

\bibitem{LaparoscopicSynthesis2019}
Pfeiffer, M., Funke, I., Robu, M.R., et~al.: Generating large labeled data sets
  for laparoscopic image processing tasks using unpaired image-to-image
  translation. In: Shen, D., et~al. (eds.) Medical Image Computing and Computer
  Assisted Intervention -- MICCAI 2019. pp. 119--127. Springer International
  Publishing, Cham (2019)

\bibitem{SCD}
Radau, P., Lu, Y., Connelly, K., Paul, G., Dick, A., Wright, G.: Evaluation
  framework for algorithms segmenting short axis cardiac {MRI}.  (07 2009)

\bibitem{Ronneberger}
Ronneberger, O., Fischer, P., Brox, T.: U-net: Convolutional networks for
  biomedical image segmentation. In: Navab, N., Hornegger, J., Wells, W.M.,
  Frangi, A.F. (eds.) Medical Image Computing and Computer-Assisted
  Intervention -- MICCAI 2015. pp. 234--241. Springer International Publishing,
  Cham (2015)

\bibitem{segars20104d}
Segars, W., Sturgeon, G., Mendonca, S., Grimes, J., Tsui, B.M.: {4D XCAT}
  phantom for multimodality imaging research. Medical physics  \textbf{37}(9),
  4902--4915 (2010)

\bibitem{CT_aug_SPIE2019}
Tang, Y.B., Oh, S., Tang, Y.X., Xiao, J., Summers, R.M.: {CT-realistic data
  augmentation using generative adversarial network for robust lymph node
  segmentation}. In: Mori, K., Hahn, H.K. (eds.) Medical Imaging 2019:
  Computer-Aided Diagnosis. vol. 10950, pp. 976 -- 981. International Society
  for Optics and Photonics, SPIE (2019)

\bibitem{wang2018pix2pixHD}
Wang, T.C., Liu, M.Y., Zhu, J.Y., Tao, A., Kautz, J., Catanzaro, B.:
  High-resolution image synthesis and semantic manipulation with conditional
  {GANs}. In: 2018 IEEE/CVF Conference on Computer Vision and Pattern
  Recognition. pp. 8798--8807 (June 2018)

\bibitem{wissmann2014mrxcat}
Wissmann, L., Santelli, C., Segars, W.P., Kozerke, S.: {MRXCAT}: Realistic
  numerical phantoms for cardiovascular magnetic resonance. Journal of
  Cardiovascular Magnetic Resonance  \textbf{16}(1), ~63 (2014)

\bibitem{ACE2019}
Wu, Z., Wang, X., Gonzalez, J.E., Goldstein, T., Davis, L.S.: {ACE:} adapting
  to changing environments for semantic segmentation. CoRR
  \textbf{abs/1904.06268} (2019)

\bibitem{2018Yasaska}
Yasaka, K., Abe, O.: Deep learning and artificial intelligence in radiology:
  Current applications and future directions. PLOS Medicine  \textbf{15}(11),
  ~1--4 (11 2018)

\bibitem{ReviewGAN}
Yi, X., Walia, E., Babyn, P.: Generative adversarial network in medical
  imaging: A review. Medical Image Analysis  \textbf{58},  101552 (2019)

\end{thebibliography}

\newpage

\begin{subappendices}
\begin{center}
\textbf{\large {--}Supplemental Document{--}}
\end{center}
\renewcommand{\thesection}{\Alph{section}}%
\section{Dataset Details}
\label{sec:DatasetDetails}
Table~\ref{tab:data} presents an overview of the number of available images, existing annotations and the application of each dataset used in this paper. MYO, LV and RV stand for left ventricular myocardium, left ventricle blood pool and right ventricle blood pool respectively.
\begin{table}[htbp!]
\centering
\caption{ A summary of different datasets used in this paper.}
\label{tab:data}
\resizebox{\textwidth}{!}{%
\begin{tabular}{llll}
\hline 
\textbf{Dataset} &
  \textbf{Number of images} &
  \textbf{Labels} &
  \textbf{Application} \\ 
  \hline \hline
XCAT Simulated \cite{ISMRM2020SimXCAT} & \begin{tabular}[c]{@{}l@{}}
  66 short-axis \\ Cine MR images \end{tabular} & 
  \begin{tabular}[c]{@{}l@{}}MYO, LV, RV, \\
  lung, liver, body tissue,\\ abdominal organ, background\end{tabular} &
  \begin{tabular}[c]{@{}l@{}}Training network 1 \\  Inference, XCAT-GAN\end{tabular} \\ \hline
ACDC \cite{ACDC} &
\begin{tabular}[c]{@{}l@{}}
  200 short-axis \\ Cine MR images \end{tabular} 
  &
  \begin{tabular}[c]{@{}l@{}} MYO, LV, RV, \\ background\end{tabular} &
  \begin{tabular}[c]{@{}l@{}}Training, network 2 \\    Inference, network 1   \\   Inference, network 3 \end{tabular} \\ \hline
SCD \cite{SCD} &
  \begin{tabular}[c]{@{}l@{}} 45 short-axis Cine MR \\images, one slice per each \end{tabular} &
  - &
  \begin{tabular}[c]{@{}l@{}}Inference, network 2 \end{tabular} \\ \hline
York \cite{York} &
  \begin{tabular}[c]{@{}l@{}} 33 short-axis Cine MR \\ images, one slice per
  each\end{tabular} &
  - &
  \begin{tabular}[c]{@{}l@{}}Inference, network 2 \end{tabular} \\ \hline
Clinical CMR (cCMR)  &
  \begin{tabular}[c]{@{}l@{}} 156 short-axis\\ Cine MR images \end{tabular} &
  \begin{tabular}[c]{@{}l@{}} MYO, LV, RV,\\ background \end{tabular} 
  & Network 3 \\ \hline
\end{tabular}%
}
\end{table}
\section{Training Details}
\label{sec:training_details}

\textbf{U-Nets for Multi-Tissue and Cardiac Cavity Segmentation}
Architecture design and training details of networks 1 and 3 can be observed in Table~\ref{tab:networks}.\begin{table}[t]
\centering
\caption{ Training and design parameters of networks for multi-tissue and cardiac cavity segmentation.}
\label{tab:networks}
\resizebox{\textwidth}{!}{
\begin{tabular}{llll}
\hline 
\textbf{Parameter} &
 \textbf{Network 1} &
  \textbf{Network 3} \\ 
  \hline \hline
Architecture &
\begin{tabular}[c]{@{}l@{}}
  As in \cite{Ronneberger} with leaky ReLU, \\dropout regularization \\ (dropout rate = 0.5) and \\ batch normalization \end{tabular} 
  &
  \begin{tabular}[c]{@{}l@{}} As in \cite{Isensee2019nnUNetBT} with dropout \\ regularization  (dropout rate of 0.2) \\ and  instance normalization  \end{tabular} &
  \\ \hline
Train parameters  &
  \begin{tabular}[c]{@{}l@{}} Batch size of 10, \\ trained for 200 epochs \end{tabular} & \begin{tabular}[c]{@{}l@{}}
  Batch size of 32,\\ trained for a maximum of \\ 500 epochs with early stopping \end{tabular} & 
  \\ \hline
  Loss function  &
  \begin{tabular}[c]{@{}l@{}} Sum of cross-entropy and dice loss, \\ initial learning rate of $10^{-4}$ \\ and $l_2$weight decay of $5*10^{-5}$ \end{tabular} &
  \begin{tabular}[c]{@{}l@{}}
  Sum of cross-entropy and dice loss, \\ initial learning rate of $5*10^{-4}$ \\ and $l_2$weight decay of $5*10^{-5}$ \end{tabular} & \\ \hline
Training data &
  \begin{tabular}[c]{@{}l@{}} XCAT Simulated \cite{ISMRM2020SimXCAT}\end{tabular} &
 ACDC \cite{ACDC}, clinical CMR (cCMR) and synthetic XCAT  \\ \hline

\end{tabular}}
\end{table} Changes introduced in the network architecture of network 1 include the use of leaky ReLU, dropout regularization (dropout rate of 0.5) and batch normalization as a regularization step. On the other hand, the best performance for network 3 is achieved by utilizing dropout regularization with the dropout rate of 0.2 and instance normalization instead of batch normalization. Random scaling and rotations, elastic deformation and mirroring are applied on the fly during the training procedure of network 1 as a data augmentation step. 
Both networks 1 and 3 are trained using the sum of cross-entropy and dice loss as a loss function and optimized using the Adam optimizer for stochastic gradient descent. Batch sizes of 10 and 32 and initial learning rates of $10^{-4}$ and $5*10^{-4}$ are utilized for networks 1 and 3, respectively, with the same weight decay $l_2$ of $5 * 10^{-5}$. Moreover, in network 3 we decrease the learning rate by a factor of 0.2 each time the exponential moving average of the training loss plateaus within the last 30 epochs. We apply early stopping when the learning rate drops below $10^{-6}$ or exceeds the maximum of 500 epochs for network 3, while network 1 is trained for 200 epochs, determined experimentally as convergence is reached. The same data augmentation methods are applied in network 3 as in network 1 with the addition of gamma correction and cropping to a region of nonzero values.

\textbf{XCAT-GAN for Conditional Image Synthesis}
The network is trained using the Adam optimizer with learning rate if 0.0002, batch size of 20 on 4 NVIDIA TITAN Xp GPUs. We increased the weight of  Kullback-Leibler divergence loss~\cite{VAEsWelling} to 0.5, but the rest of the training parameters, the architectures of the encoder, generator and discriminator are the same as~\cite{park2019SPADE}. Moreover, random affine and elastic transformations are used during training. 

\end{subappendices}

\end{document}